\documentclass[usenatbib]{mn2e}

\usepackage{epsfig}
\usepackage{graphicx}
\usepackage{subfigure}

\begin{document}
\bibliographystyle{mn2e}

\newcommand{\DeltaD}{{$\Delta D$}}
\newcommand{\DeltaV}{{$\Delta V$}}
\newcommand{\ewha}{${\rm EW(H\alpha)}$}
\newcommand{\kms}{km s$^{-1}$}
\newcommand{\halpha}{H$\alpha$}

\title[Exploring the Links between Star Formation and Minor Companions around Isolated Galaxies]{Exploring the links between star formation and minor companions around isolated galaxies}
\author[\sc Jacob Edman, Elizabeth J. Barton, and James S. Bullock]{Jacob P. Edman, Elizabeth J. Barton, and James S. Bullock$^{1}$\\
$^{1}$Center for Cosmology, Department of Physics and
Astronomy, University of California, Irvine,  CA 92697-4575
(email: ebarton@uci.edu)}

\date{9 October 2011}

\maketitle

\begin{abstract}

Previous studies have shown that galaxies with minor companions
exhibit an elevated star formation rate.  We reverse this inquiry,
constructing a volume-limited sample of $\sim$L$^{\star}$ (M$_{\rm r}
\leq -19.5+5\log{h}$) galaxies from the Sloan Digital Sky Survey that
are isolated with respect to other luminous galaxies. Cosmological simulations suggest
that 99.8\% of these galaxies are alone in their dark matter haloes with respect to other luminous 
galaxies. We search the
area around these galaxies for photometric companions.  Matching
strongly star forming (\ewha $\geq 35$~\AA) and quiescent
(\ewha $ < 35$~\AA) samples for stellar mass and redshift using a Monte
Carlo resampling technique, we demonstrate that rapidly star-forming
galaxies are more likely to have photometric companions than other
galaxies.  The effect is relatively small; about 11 \% of quiescent, isolated
galaxies have minor photometric companions at radii $\leq 60$ kpc h$^{-1}$
kpc while about 16 \% of strongly star-forming ones do. Though small, the cumulative difference
in satellite counts between strongly star-forming and quiescent galaxies is highly
statistically significant (P$_{KS}$ = 1.350 $\times 10^{-3}$) out to to radii of $\sim
100$ h$^{-1}$ kpc.  We discuss explanations for this excess, including the
possibility that $\sim 5 \%$ of strongly star-forming galaxies have star formation that is causally related to
the presence of a minor companion. 

\end{abstract}


\section{Introduction} \label{sec:intro}

Many studies have established that the environmental properties
of galaxies are closely related to their star formation histories \citep[e.g.,][]{Dressler1980,Postman1984}.  On
large scales, stronger clustering effects are observed in red galaxies
than in blue galaxies, because red galaxies also tend to be more
massive and reside in more massive dark matter halos \citep[see][and references therein]{2011Zehavi}.
However, at scales less than $\sim 300$ kpc, where
galaxies are likely to be part of the same halo, these clustering
effects are poorly understood, and at scales less than 50 kpc, blue
galaxies may cluster more strongly \citep{2008ApJ...679..260M}.

Previous studies establish that galaxy interactions
and the presence of a companion are both associated with an increased
rate of star formation \citep[e.g.]{barton,
lambas,LarsonTinsley,Mihos1994}.  In particular, closer companions lead to increased rate of
star formation for both major and minor interactions \citep[e.g.]{barton,Lambas2003,Nikolic2004,Woods2006}.
These results are qualitatively in line with expectations from numerical simulations  that predict enhanced star formation in interacting pairs before the galaxies finally merge 
 \citep[e.g.]{Barnes1996,Mihos1994}.
 though those results are sensitive to uncertain star formation physics (e.g. \citep{2006Cox}). In principle, close passes in both major and minor pairs can initiate a burst of star formation.

The questions that surround minor mergers are closely related to the
properties of the satellite galaxies and their hosts.  The radial distribution of
satellites around primary galaxies has been extensively explored.
Chen (2008) finds that the radial distribution of blue satellites around isolated hosts is significantly shallower than that of red satellites.  There were also hints in this study that red and blue hosts   may have different satellite distributions
\citep{2008A&A...484..347C}.  Previous studies have also examined the
effects of galaxy color on the anistropic distribution of satellite
galaxies around their hosts \citep{2007AAS...211.0901A,
  2007MNRAS.376L..43A, 2008MNRAS.390.1133B, 2007MNRAS.378.1531K}. For
red central galaxies, satellites tend to be more strongly aligned
along the major axis, while the distribution of satellites for blue
central galaxies are consistent with an isotropic distribution. These
observations further demonstrate the expectation that relationships
exist between galaxy color and satellite distribution.

Some studies reveal a deficiency of satellites at very small radii
(less than 20 kpc) compared to the extrapolated outer density profile
\citep{2005MNRAS.356.1045S, 2005MNRAS.356.1233V}. This deficiency may be due to a failure to resolve
interacting pairs into single galaxies, or it could be a real
deficiency due to galaxy destruction. Sales \& Lambas
(2005) find that this deficiency can be predictably be found in poorly
star forming galaxies, which tend to have larger core radii, while
strongly star forming galaxies often have more close neighbors than
predicted \citep{2005MNRAS.356.1045S}. This finding lends support to the long-established
hypothesis that interaction with a nearby galaxy often causes enhanced
star formation.

Despite our knowledge that both major and minor interactions
trigger star formation, the complete cosmological impact of these
processes remain unknown.  When galaxies do exhibit significantly
enhanced star formation, it is not known how often that star formation
is related to an interaction.  Here, we invert the standard approach
to minor mergers by identifing a volume-limited sample of
$\sim$L$^{\star}$ galaxies that are isolated with respect to other
luminous galaxies in from the Sloan Digital Sky Survey Data Release 6
\citep[SDSS6][]{Adelman-McCarthy2008}.  We then compare the photometric companion counts
around the star-forming galaxies to the counts around other galaxies.
In \S~2, we describe the initial galaxy sample and search procedures.
\S~3 describes the bias corrections we apply to ensure a fair
comparison.  We describe the results in \S~4, discuss them in \S~5,
and conclude in \S~6.

\section{Sample} 
We begin by identifying a set of galaxies in the Value Added Galaxy 
Catalog of
SDSS DR6  \citep[VACG]{Adelman-McCarthy2008},\citep{Blanton2005}.
We restrict the sample to sources that are
isolated with respect to other luminous galaxies.  We compile all
sources in the Sloan Digital Sky Survey Data Release 6 that are more
luminous than M$_{\rm r}$+ 5 log (h)= -19.5 at a redshift $z <
0.0804$, which is the upper limit at which a galaxy of this luminosity
will appear in the spectroscopic sample.  We then identify galaxies
that have no neighbors within a projected distance of 
400 kpc h$^{-1}$ and 1000 \kms of the primary, and at most one neighbor
within 700 kpc h$^{-1}$ and 1000 \kms.
These conservative isolation criteria include neighbors
with measured redshifts and potential neighbors which are in the
photometric survey but do not have measured redshifts. For example, 
a galaxy with a potential luminous neighbor with unmeasured
redshift at 300 kpc h$^{-1}$ will be rejected. 

These criteria result in a sample of 24,753 isolated central
galaxies. Using the techniques of Barton et al. (2007), we employ
cosmological simulations to demonstrate the expected purity of this
sample.  The method relies on assuming that halo circular velocities
in simulations scale monotonically with M$_{\rm R}$.  The space
density of galaxies with M$_{\rm r}+ 5 log (h)= -19.5$ in SDSS,
computed by integrating the luminosity function of \citep{Blanton2003},
corresponds to halos that had circular velocities $\geq
164$ \kms at the present epoch, or at the time they became substructure if
they are not central sources in the hybrid n-body and semi-analytic
simulations of \citep{Zentner2005}.  Forming an artificial redshift
survey with these galaxies and applying selection criteria that are
identical to those applied to the data, we show that 99.8\% of the
isolated sample consists selected via our isolation criteria galaxies
that are alone in their dark matter halos (with respect to other
galaxies with M$_{\rm r} \leq -19.5 + 5\log(h)$).

A subset of the SDSS central galaxies are near the edges of the 
survey, in less complete regions, or close to bright stars.  As a result, 
we may not be completely probing their environments for potential 
companions.  We thus identify a ``complete environment'' subsample of 
galaxies using the tools available in the VACG.  In particular, we remove 
the lowest and highest redshift sources so that we can probe the entire 
1000 \kms in front of and behind the galaxies.  Then, we use the random 
subsamples provided in the VACG, weighted by completeness of each sector 
and weighted based on the magnitude limit of the survey in that sector. 
We add the ``random counts'' in each galaxy's relevant environment, in 
essence performing a Monte Carlo integration of its environment weighted 
by completeness.  We then restrict the ``complete environment'' subsample 
to the 18,601 galaxies where the value of this integral is within 
$2\sigma$ of the mode for the entire distribution.  We explore whether 
restricting to this subsample affects any of our results.  Because it does 
not have any qualitative effect, we confine our discussion to the original 
24,853 isolated galaxy sample hereafter.

We use the SDSS Data Release 7 online database to tabulate potential minor 
companions to an angular radius that would correspond to 200 h$^{-1}$ kpc 
from each central galaxy, with an apparent magnitude limit of m$_{\rm 
R}=21$.  In the remaining sections of the paper, we explore the dependence 
of the companion counts on galaxy properties.

\begin{table}
\caption{Sample sizes for complete and corrected samples} 
\label{sample}
\begin{center}
\begin{tabular} {l | c |  c  | r}
cutoff type & cutoff & star-forming & quiescent  \\
\hline
g-r & 0.4  & 1825 & 22435 \\
\ewha & 35\AA &	1951 &	16467\\
g-r (Corrected) & 0.4 & 1388 & 7180\\
\ewha (Corrected) & 35\AA & 1566 & 7180\\
\end{tabular}
\end{center}
\end{table}

\section{Bias corrections}
In this study, we focus on the counts of potential faint companion
galaxies around otherwise isolated $\sim$L$^{\star}$ central galaxies.
In particular, we separate pair samples into rapidly star-forming
galaxies and galaxies that are not rapidly star forming.  Because we
do not have redshifts for these potential comanions, we must count
them in an angular radius around the centrals.  In addition, because
we count them to a fixed flux limit, the study will include relatively
less luminous companions around nearby galaxies.  As a result, any
differences in the redshift distributions of two samples of galaxies
that we compare will cause potential biases in the comparisons.
Because galaxies are known to cluster, and because satellites should
be more numberous around more massive galaxies, halo mass is another
key parameter that must be controlled when comparing two samples.
No direct measure of halo mass is available, so we use stellar
mass as a proxy.  In this section, we describe our simple techniques
to resample the data in order to compare galaxies with the same
distributions of redshift and stellar mass.

We employ a Monte-Carlo random selection method to compare the
radial distributions of neighbors around blue and red central
galaxies with similar redshifts and stellar masses. Our Monte-Carlo simulation selects a
random sample of 5 red galaxies with similar redshift and
stellar mass to each of the blue galaxies in the sample. A red galaxy
is considered similar to the blue galaxy if z is within $\pm$ 0.005
and the stellar mass is within $\pm$ 10\%.

\begin{figure}
\includegraphics[width=85mm]{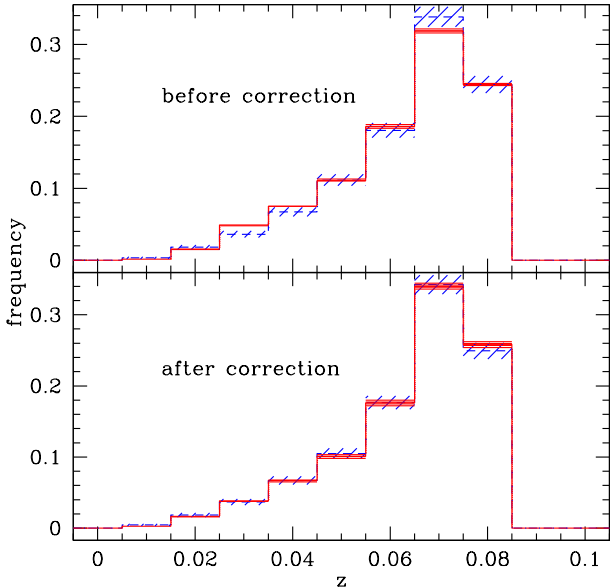}
\caption{\label{redshift}\emph{Top:} The redshift distribution of central galaxies for the complete sample of galaxies. \emph{Bottom:} The redshift distribution of central galaxies is shown for a corrected sample. For both plots, the blue dashed line depicts galaxies with \ewha $\ge$ 39 \AA\ and the red solid line shows galaxies with \ewha < 35 \AA. 
Error bars are calculated for each bin as $\sqrt{n}/n_{total}$.}
\end{figure}

\begin{figure}
\includegraphics[width=85mm]{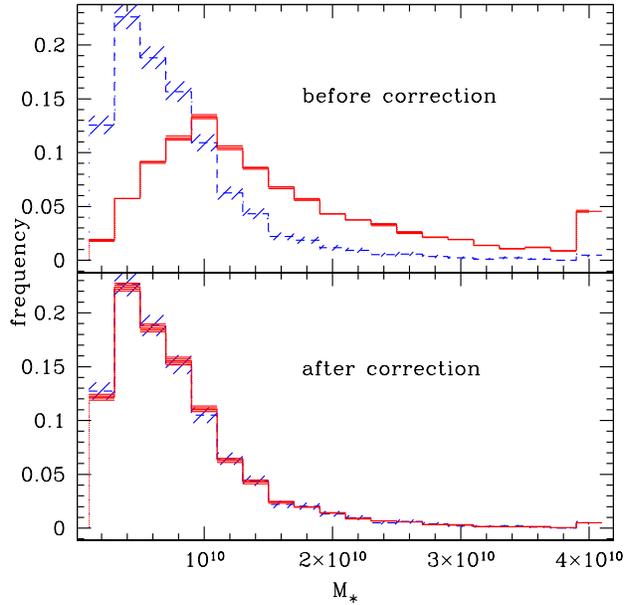}
\caption{\label{stellarmass}\emph{Top:} The stellar mass  distribution of central galaxies for the complete sample of galaxies. \emph{Bottom:} The stellar mass  distribution of central galaxies is shown for a corrected sample. For both plots, the blue dashed line depicts galaxies with \ewha $\ge$ 35 \AA\ and the red solid line shows galaxies with \ewha < 35 \AA.
Error bars calculated for each bin as  $\sqrt{n}/n_{total}$.}
\end{figure}

Figure \ref{redshift} depicts the redshift distributions of strongly star-forming and quiescent central galaxies, 
based on \ewha~ cutoff of 35 \AA, 
for the complete sample and one of the subsamples derived using our Monte-Carlo selection method. 
Visually, the redshift distributions of the strongly star-forming and quiescent  galaxies in the complete sample are significantly different.  There is an excess of strongly star-forming galaxies  at lower redshifts when compared 
to the quiescent galaxies. The corrected sample appears to match much more closely.  
This interpretation is supported by the
Kolmogorov-Smirnov test results, which are reported in Table \ref{ks}. 
P$_{KS}$ (z) = 5.417 $\times$ 10$^{-9}$ for the full sample, which indicates that the redshifts of star-forming and quiescent galaxies have a different distribution. For the corrected sample, P$_{KS}$ (z) = 0.5746, so the two subsamples likely are 
drawn from the same distribution. 

The stellar mass distributions for the complete sample, plotted in the top of
Figure \ref{stellarmass}, are even more disparate than the redshift distributions. As
expected, there is an excess of centrals with \ewha $<$ 35 \AA\ in the
high-mass range, and an excess of centrals with \ewha $>$ 35 \AA\ in
the low mass range. The extremely low P$_{KS}$ (z) and P$_{KS}$ (stellar mass)
values for the subsamples taken from the complete set of
central galaxies demonstrate that they have
extremely different redshift and stellar mass distributions.
The Monte-Carlo derived sample, plotted at the bottom of Figure \ref{stellarmass}, provides a much better visual match between the stellar mass distributions of strongly and not strongly star-forming central galaxies. Once again, this interpretation is supported by the KS test results, which are shown in Table \ref{ks}.
We show the KS statistics from 10 Monte-Carlo simulations, and KS (z)
ranges from 0.249 - 0.696 and the KS (stellar mass) spans
0.818-0.969.

\begin{table} 
\caption{\label{ks}Results of Kolmogorov-Smirnov test for redshift and stellar mass distributions of \ewha separated central galaxies and Monte-Carlo samples} 
\begin{center} 
\begin{tabular}{l | c | c |  c | } 
Sample type & cutoff & P$_{KS}$ (z) & P$_{KS}$ (stellar mass)  \\
\hline 
Complete & \ewha = 35 \AA & 0.01688 & 0 \\
Complete & g-r = 0.4 & 5.417 $\times 10^{-9}$ & 0\\
Corrected & \ewha = 35 \AA & 0.249 - 0.696 & 0.818 - 0.969 \\
\end {tabular} 
\end{center} 
\end{table}

Additionally, because we necessarily observe a two-dimensional view of astronomical
objects, our sample of potential satellites is
contaminated by interlopers, i.e. objects that are not physically within
our searched radius but nevertheless appear to be true companions when
projected on the sky. Overall, interlopers tend to flatten the
projected radial mass distribution, and the fraction of observed
satellites that are actually interlopers increases with search
radius. \citep{2006ApJ...647...86C} tested both volume limited and flux
limited samples (which are biased towards brighter satellites) and
found consistent radial distributions, showing that there is at best a
limited dependence of interloper contamination on magnitude of the
satellites and color of the primaries. Sales and Lambas (2005) find a flat
distribution of interlopers between 20 and 500 kpc by specifically
sampling for companion galaxies with a large projected velocity
difference ($2000 < |\Delta V| < 10000$ \kms) compared to the central,
allowing them to assume a uniform contamination when calculating the
radial density profiles of satellites.

For our study, interloper contamination will obscure the
relationship between close satellites and blue hosts, because red
galaxies tend to be more massive and reside in larger dark matter
halos than blue galaxies. This larger halo size means there is a
stronger clustering effect on scales of $\sim 300$ kpc and larger
\citep{2008ApJ...679..260M}. When searching on small radial scales (for
example, less than 100 kpc) there should be minimal contamination, but
potential satellites at larger radii in these larger dark matter halos
are more likely to be interlopers because the environment is more
crowded. 
Assuming a isotropic background, if interlopers are a significant factor in our study,
we would expect the radial distribution of 
satellites to be roughly proportional to r$^{2}$. However, as we shall demonstrate below, the radial distribution we observe scales closer to $\sim$r. At small radii, the true companions appear to dominate.  

\section{Results}

We compare the list of potential companions with their associated
central galaxies and compute the projected radial distance to each
companion, throwing out any companions at a distance that exceeds the
specified search radius, or that have a Petrosian r magnitude greater than 21. For our
primary analysis, we choose to set the red-blue cutoff for our central
galaxies at \ewha = 35 \AA . This cutoff allows us to examine only the
most strongly star forming galaxies, giving us the best chance of
detecting the effect of close neighbors on galaxy star formation
activity.

\subsection{Numbers of neighbors}

\begin{figure}
\subfigure[]{
\includegraphics[width=85mm]{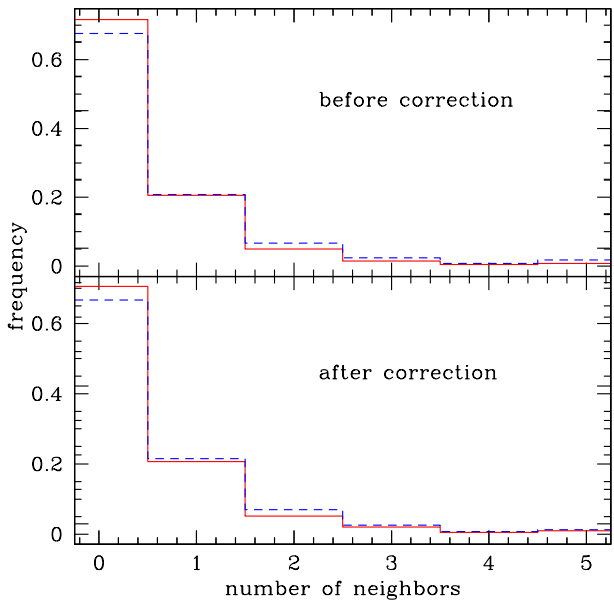}
\label{n50}
}
\subfigure[]{
\includegraphics[width=85mm]{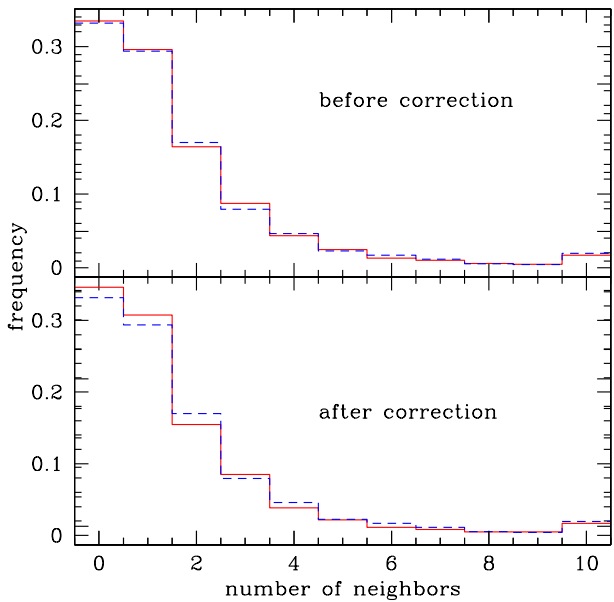}
\label{n200}
}
\label{histo}
\caption{Histograms of the number of neighbors for each central galaxy within 100 kpc (top) and 200 kpc (bottom). Within each plot, the top plot is the complete sample and the bottom is the corrected sample. The blue dashed line indicates central galaxies with \ewha $\ge$ 35 \AA; the red solid line indicates central galaxies with \ewha < 35 \AA.}
\end{figure}

In figures \ref{n50} and \ref{n200}, we demonstrate that strongly star-forming central
galaxies are more likely to have a close neighbor. Figure \ref{n50} contains
histograms of the frequency of a central galaxy having a given number
of neighbors within 100 kpc for the complete (top) and corrected (bottom) samples. We choose
to plot 100 kpc because closer neighbors are associated with more
recent starbursts, and thus larger stellar populations in the blue
spectral range \citep{barton}. Again, the central galaxies are grouped into strongly star-forming
and quiescent subsamples, with the cutoff at \ewha =35 \AA. The quiescent
central galaxies are represented by the solid red line and the strongly star-forming
centrals by the dashed blue line. The frequency is expressed as a
fraction of the total neighbors in each category.

For both the complete and the corrected sample, there 
is an excess of strongly star-forming central galaxies with one or two neighbors within 100 kpc. 
We find that the KS probability that the neighbor distributions are the same is zero,
indicating an extremely high degree of significance. 
For greater numbers of neighbors, the red and blue lines are nearly identical. Figure \ref{n200} is similar to the
figure \ref{n50} but for a radius of 200 kpc. At this radius, the
distribution of number of neighbors for both types of centrals centrals is 
more similar, but P$_{KS}$ = 0, which indicates again that the observed (small) difference has a high degree of significance. 

We made similar plots for radii ranging from 20 to 200 kpc, and a
few trends become obvious. As the search radius decreases, the
difference between star-forming and quiescent centrals becomes magnified, indicating
that there is a relationship between star formation and distance to the
nearest companion \citep{2006ApJ...647...86C}. With increasing radius, the line representing the
quiescent central galaxies becomes closer and closer to the blue line, until
they become nearly indistinguishable around 150 kpc. This behavior is
expected, because at large radii, the effect a companion can have on triggering
star formation is small. Additionally, the signal is increasingly dominated by interlopers as
the searched radius increases. Finally, star-forming galaxies
tend to have one close neighbor, while quiescent galaxies tend to have a
large number of neighbors at larger distances away. This result indicates
that some of the star-forming central galaxies may be part of an interacting
pair of galaxies, possibly triggering a burst of star formation. 

The behavior of the quiescent centrals arises because red galaxies are
more massive and thus can be expected to
cluster more strongly. For all samples at all radii,
Kolmogorov-Smirnov tests indicate a very high degree of statistical
significance ($>$ 99.999\%) for the difference between distributions of
number of neighbors, whether divided by g-r color or
\ewha. Ultimately, these plots suggest that star-forming central galaxies tend
to have more near neighbors than quiescent central galaxies, which lends
support for the idea that many are  because of a triggered star
formation event.

\subsection{Radial distribution of neighbors}

\begin{figure}
\subfigure[]{
\includegraphics[width=85mm]{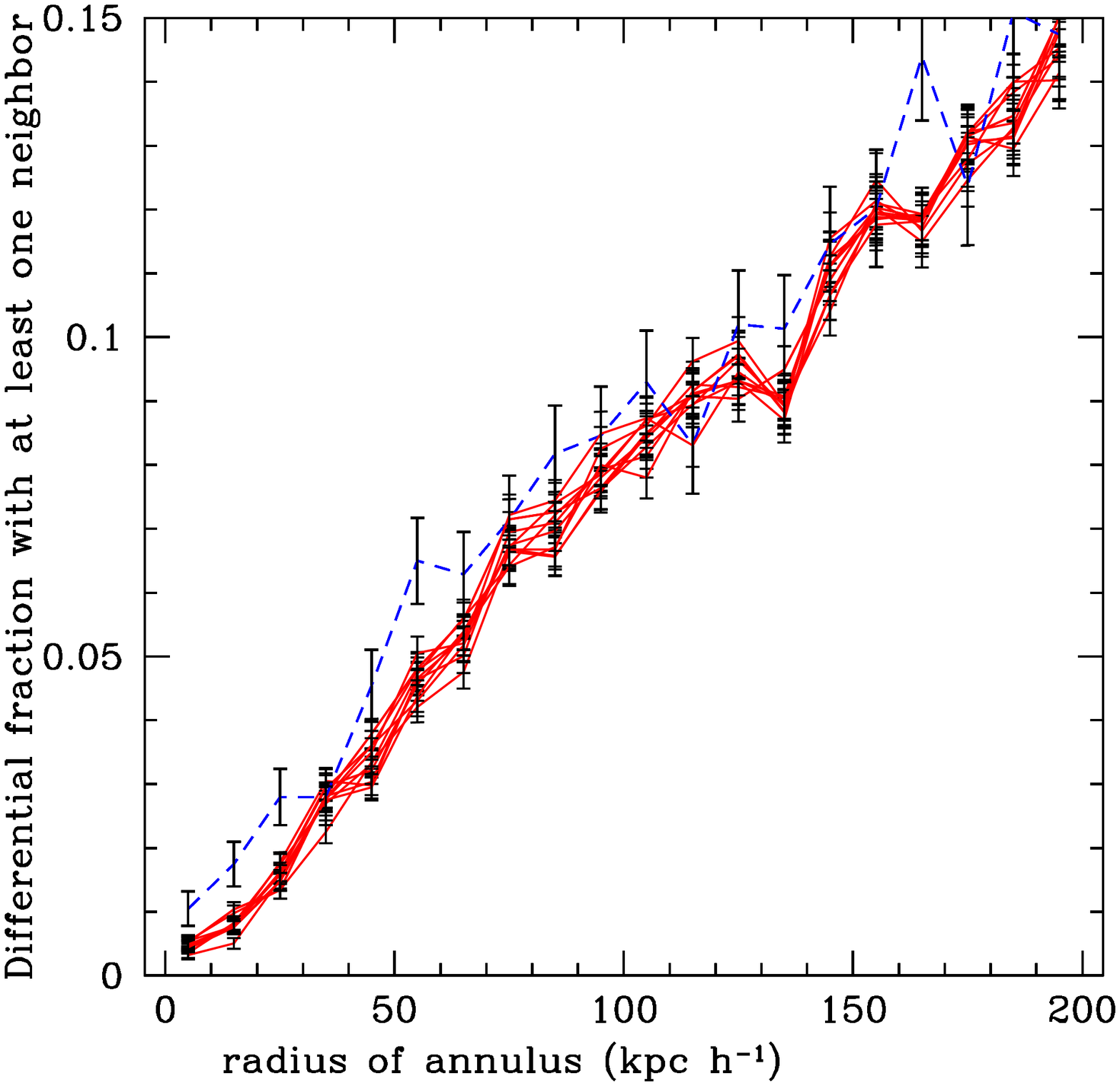}
\label{annulus}
}
\subfigure[]{
\includegraphics[width=85mm]{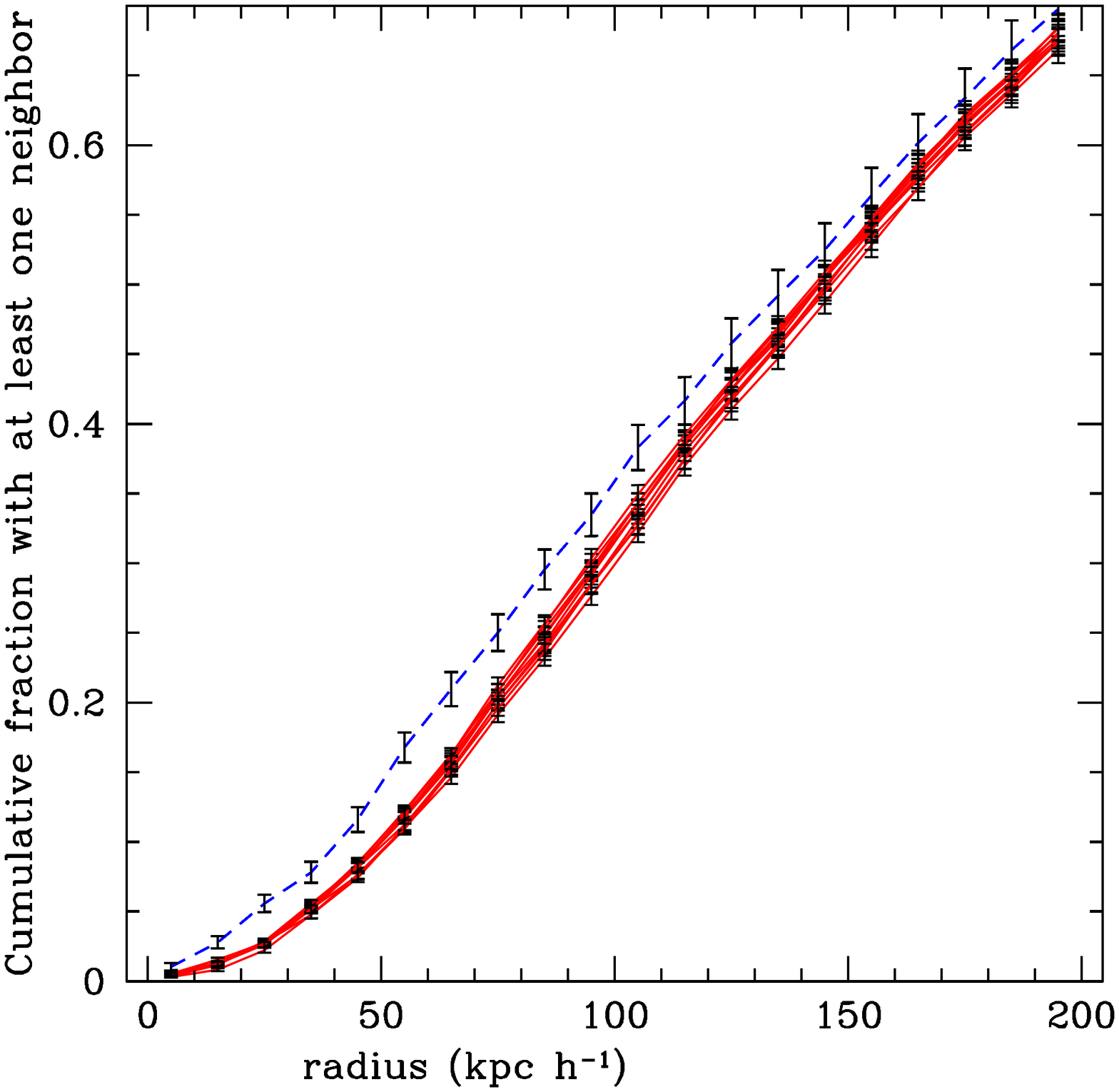}
\label{radius}
}
\caption{\emph{Top}: The fraction of central galaxies with at least one neighbor in decadal bins ranging from 0 - 200 kpc for 10 Monte-Carlo selected samples. The blue dashed line indicates strongly star-forming galaxies with \ewha $\ge$ 35 \AA, while the red solid lines are centrals with \ewha $<$ 35 \AA.  \emph:{Bottom}: Similar to \emph{Top}, but here we plot the cumulative fraction of centrals with at least one neighbor within a given radius for 10 Monte-Carlo samples.}
\label{radial}
\end{figure}

After showing that central galaxies above the \ewha\ cutoff are
somewhat more likely to have a near neighbor than centrals below the
cutoff, we quantify difference in radial distribution of
satellites between the strongly star-forming and less star-forming centrals.
For this, we use the redshift and stellar mass corrected Monte-Carlo samples. Each
Monte-Carlo sample chooses 5 red central galaxies similar to each blue
central. In figure \ref{radial}, we plot the radial distribution of
neighbors for 10 of these Monte-Carlo samples in two different ways.

Figure \ref{annulus} shows the fraction of central galaxies with at least one
neighbor in each annulus, with annulus bins in steps of 10 kpc. The
blue dashed line represents centrals with \ewha $>$ \AA\ and the red
lines are the quiescent centrals with \ewha $\le$ 35\AA. There is some variation
of the quiescent subsamples because of the random selection, but it is
well within the Poisson error bars shown. Inside 
85 kpc, central galaxies above the \ewha break
are significantly more likely to have a neighbor. This result clearly
indicates that the strongly star-forming galaxies are more likely to have a neighbor
nearby. Beyond 85 kpc, the plot gets noiser, and there is little
significant difference between the red and blue lines. We expect this
because at such large radii a neighboring galaxy can have little
direct effect on the star formation of the central galaxy, and interlopers begin to dominate \citep{Tollerud2011}.

Figure \ref{radius} is similar to figure \ref{annulus}, but instead we
plot the cumulative fraction of central galaxies with at least one neighbor inside a given
radius. At the innermost bin, centered at 5 kpc, the gap narrows; objects separated
by less than 3 kpc (including mergers, which are often associated with
bursts of star formation) are nearly impossible for the detector to
resolve.  We find that the cumulative difference between star-forming and quiescent centrals is
significant out to 100 kpc, implying that star-forming galaxies are $\sim 5 \%$ more
likely to have at least one neighbor within 100 kpc than quescent galaxies of the same stellar mass.
However, the overall fraction of star-forming
galaxies with a neighbor within 100 kpc is still small.

\begin{table} 
\caption{\label{ks_nn}Results of Kolmogorov-Smirnov test for nearest neighbor distributions of central galaxies} 
\begin{center} 
\begin{tabular}{l | c | c | } 
	Sample type & comparison cutoff & P$_{KS}$ (neighbor)  \\
\hline
\hline 
Complete & \ewha = 35 \AA & 0.02642 \\
Complete & g-r = 0.4 & .01445 \\
\hline
Corrected & \ewha = 20 \AA & 8.452 $\times 10^{-2}$\\
Corrected & \ewha = 35 \AA & 1.350 $\times 10^{-3}$\\
Corrected & \ewha = 40 \AA & 8.615 $\times 10^{-3}$\\
Corrected & g-r = 0.4 & 0.1390 \\
Corrected  & g-r = 0.5 & 0.1022 \\
Corrected  & g-r = 0.6 & 2.269 $\times 10^{-3}$ \\
\end {tabular} 
\end{center}
\end {table}

The difference in nearest neighbor distribution between strongly star-forming 
and quiescent galaxies is significant as demonstrated by the KS statistics,
which are tabulated in Table \ref{ks_nn}.   Here we present comparisons for several choices for how to
divide the populations between quiescent or red galaxies and star-forming or blue galaxies, as specified by a "cutoff" criterion in the middle column,
either in g-r color or \ewha.
The Corrected samples used in this analysis contain 20 
quiescent  or red central galaxies for each strongly star-forming or blue central galaxy.

The Complete sample statistics, presented in the top two rows, include the entire sample of 24,753 galaxies, 
uncorrected for redshift or stellar mass.  We find the neighbor distributions within the Complete sample to be marginally different for cutoffs at both g-r = 0.4 and 
\ewha = 35 \AA, with P$_{KS}$ values of 0.01445 and 0.02646, respectively. 
The Corrected samples demonstrate more significant differences.  
The highest degree of significance is found for the cutoff at \ewha = 35 \AA, which has P$_{KS} = 1.350 \times 10^{-3}$.
However, when our resampling method is applied using g-r color as the cutoff, the difference between the nearest
neighbor distributions for blue and red central galaxies becomes less significant, possibly due to the differing timescales of g-r as an indicator of recent star formation or simply a consequence of the reduced sample size. Using g-r color as a proxy for star formation is also prone to errors due to dust-reddening, which would tend to cancel out the effect we see when using \ewha.

\begin{figure}
	\includegraphics[width=85mm]{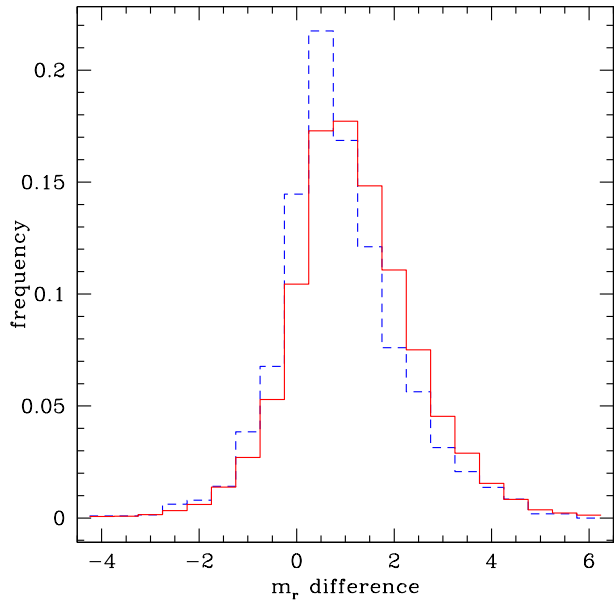}
	\caption{Histogram of the differences in apparent r-band magnitude for each companion and its host. The blue dashed line is for companions of centrals with \ewha $\ge$ 35 \AA\, and the red solid line is companions of centrals with \ewha $<$ 35 \AA.}
	\label{mag}
\end{figure}

Figure \ref{mag} is a histogram of the differences in apparent r-band magnitude for each companion and its host. The blue dashed line represents companions of strongly star-forming hosts, while the red solid line is quiescent hosts. From this plot, we see that strongly star-forming central galaxies tend to have relatively brighter companions than quiescent centrals.  

\section{Discussion}
In summary, we selected a sample of isolated galaxies from the Sloan Digital Sky Survey, and by employing a Monte-Carlo resampling technique, demonstrated that strongly star-forming and quiescent galaxies exhibit different probabilities of having a
close companion, even when samples are selected to have matching redshift and stellar mass distributions. When using \ewha~ as a proxy for star formation, these differences are small (star-formers are $\sim 5$\% more likely to have a close neighbor) but highly significant (P$_{KS}=\sim 10^{-3}$). 
Thus, we have shown that strongly star-forming and quiescent galaxies exhibit different
radial distributions of photometric companions. 

\begin{figure}[]
		\includegraphics[width=85mm]{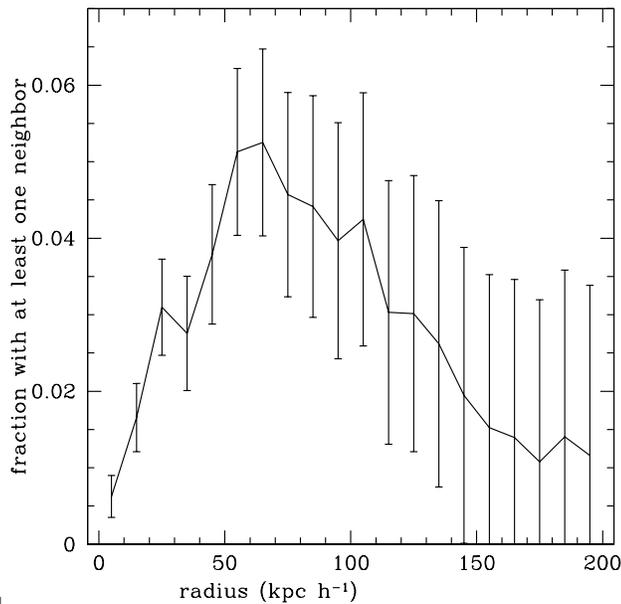}
	\caption{The difference in the cumulative fraction of strongly star-forming and quiescent central galaxies with a neighbor within a given radius. Strongly star-forming central galaxies have \ewha $\ge$ 35 \AA. }
	\label{diff}
\end{figure}

We summarize our primary result in Figure \ref{diff}, which is similar to Figure \ref{radius}, but here we plot the difference in the cumulative fraction of
strongly star-forming and quiescent central galaxies as a function of separation.  We are again using  \ewha\ = 35 \AA~ as the dividing criterion to separate star-forming and quiescent galaxies. 
The maximum difference is $5 \pm$ 2\% in the 55 kpc radius bin. We explored the effects of using different \ewha cutoffs in our simulations, and the results are described in \ref{ks_nn}.

There are several possible explanations for the 5\% excess in companions seen for the strongly star-forming central galaxies.
One possibility is that galaxies are more likely to be star forming if they have a close companion -- that is, some small fraction
of these star forming galaxies ($\sim 5\%$) are star forming {\em because} they are experiencing an  
interaction with a close companion.  This is a mechanism that has been well-documented in previous 
studies \citep{LarsonTinsley,barton,lambas,Woods2006,Mihos1994}.  Alternatively, there is something else about having an elevated star formation rate
that increases a galaxy's likelihood to have a close companion.  One possibility is that objects that are currently accreting gas via ``cold mode" 
deposition are more likely to have companionship, as this source of fuel is usually associated with larger-scale filamentary over-densities in galaxy positions \citep[see][and references therein]{2009Keres}.  

Finally, it is possible that there is some systematic error in the determination of M$_{\ast}$ which would obscure the fact
that the strongly star-forming central galaxies in our sample reside in more massive haloes than the less
strongly star-forming central galaxies. However, this is unlikely, because red galaxies (which generally have lower star formation rates) tend to be in higher mass haloes than blue galaxies.

\bibliography{astro,astro2}

\end{document}